\documentclass[twocolumn,nofootinbib,amsmath,amssymb,aps,superscriptaddress,pre]{revtex4-1}
\usepackage[pdftex]{graphicx}% Include figure files
\usepackage{amssymb,amsfonts,amsmath}
\usepackage{color}
\usepackage[usenames,dvipsnames]{xcolor}
\definecolor{lightblue}{rgb}{0.88, 0.90, 1.0}
\definecolor{darkchocolate}{rgb}{0.55, 0.27, 0.07}

\begin{document}
\title{Do thermodynamically stable rigid solids exist?}
\author{Parswa Nath}
\affiliation{TIFR Centre for Interdisciplinary Sciences, 36/P Gopanpally, Hyderabad 500107, India}
\author{Saswati Ganguly}
\affiliation{Institut f\"ur Theoretische Physik II: Weiche Materie, Heinrich 
Heine-Universit\"at D\"usseldorf, Universit\"atsstra{\ss}e 1, 40225 D\"usseldorf, 
Germany}
\author{J\"urgen Horbach}
\affiliation{Institut f\"ur Theoretische Physik II: Weiche Materie, Heinrich 
Heine-Universit\"at D\"usseldorf, Universit\"atsstra{\ss}e 1, 40225 D\"usseldorf, 
Germany}
\author{Peter Sollich}
\affiliation{King's College London, Department of Mathematics, Strand, London WC2R 2LS, U.K.}
\author{Smarajit Karmakar}
\affiliation{TIFR Centre for Interdisciplinary Sciences, 36/P Gopanpally, Hyderabad 500107, India}
\author{Surajit Sengupta}
\affiliation{TIFR Centre for Interdisciplinary Sciences, 36/P Gopanpally, Hyderabad 500107, India}

\date{\today}
\begin{abstract}
Customarily, crystalline solids are defined to be {\em rigid} since they resist changes of shape determined by their boundaries.  However, rigid solids cannot exist in the thermodynamic limit where boundaries become irrelevant. Particles in the solid may rearrange to adjust to shape changes eliminating stress without destroying crystalline order. Rigidity is therefore valid only in the {\em metastable} state that emerges because these particle rearrangements in response to a deformation, or strain, are associated with slow collective processes. Here, we show that a thermodynamic collective variable may be used to quantify particle rearrangements that occur as a solid is deformed at zero strain rate. Advanced Monte Carlo simulation techniques are then employed to obtain the equilibrium free energy as a function of this variable. Our results lead to a new view on rigidity: While at zero strain a rigid crystal coexists with one that responds to infinitesimal strain by rearranging particles and expelling stress, at finite strain the rigid crystal is metastable, associated with a free energy barrier that decreases with increasing strain. The rigid phase becomes thermodynamically stable by switching on an external field, which penalises particle rearrangements. This produces a line of first-order phase transitions in the field - strain plane that intersects the origin. Failure of a solid once strained beyond its elastic limit is associated with kinetic decay processes of the metastable rigid crystal deformed with a finite strain rate. These processes can be understood in quantitative detail using our computed phase diagram as reference.
\end{abstract}
%
%\verticaladjustment{-2pt}
\maketitle
%\thispagestyle{firststyle}
%\ifthenelse{\boolean{shortarticle}}{\ifthenelse{\boolean{singlecolumn}}{\abscontentformatted}{\abscontent}}{}

\section{Introduction}
The ability to resist changes of shape, or rigidity, has
been explained as a consequence of the spontaneous breaking of
continuous translational symmetry in crystalline solids~\cite{CL,szamel1}.
Quite surprisingly, this result is at the same time, paradoxical.
It may be shown quite rigorously~\cite{ruelle} that any homogeneous {\em bulk} deformation created within a solid to conform to changes of shape of the boundary may always be accommodated instead by {\em surface} distortions involving particle rearrangements~\cite{penrose,sausset}. 
This automatically suggests that any internal stress generated {\em in equilibrium}
within a macroscopically large solid in response to a change of
shape must necessarily vanish~\cite{penrose}. Given enough time, a solid always flows to release this stress under any external mechanical
load, however small~\cite{sausset}. The emergence of rigid
solids is therefore associated with inherently long-lived {\em metastable} states~\cite{penrose, shibu}. This fundamental result has, quite understandably, wide ranging
connotations in the science and technology of materials, especially
with regard to properties like high temperature creep, fatigue,
fracture and plastic flow~\cite{rob,rei}.
%Contradiction can be circumvented by defining statistical mechanical
%averages over restricted ensembles~\cite{penrose,shibu} or equivalently
%over finite times~\cite{sausset} implying that all rigid solids are
%inherently long-lived {\em metastable} states.

While the immediate paradox is resolved, we still need to address the question of how a rigid crystal, when deformed, releases internal stress and transforms to a flowing state. A fundamental understanding of this process should also reveal under what conditions thermodynamically stable rigid crystals may exist.  Here, a comparison to fluids in the limit of zero strain rate is very instructive. While fluids subjected to small stresses exhibit Newtonian flow with a constant viscosity~\cite{CL}, no such regime exists for stressed solids whose viscosity diverges with vanishing
stress~\cite{sausset}. Does this singular behaviour of the viscosity imply an underlying phase transition? Moreover, distinct from the fluid state, flow in a crystal is triggered by the formation of slip planes~\cite{penrose} causing rearrangements of particle neighbourhoods. Rigid solids composed of distinguishable particles are thus also associated with the breaking of {\em discrete} permutation symmetry.  

In this work, we show, to the best of our knowledge for the first time, that a phase transition indeed occurs at zero strain rate. A static, {\em equilibrium}, first-order phase transition describes the transformation of one crystal to another with identical crystal structure but with differently arranged local neighbourhoods. Stress relaxation occurs as a consequence of these rearrangements. As expected for a first-order transition, the transformation {\em kinetics} of the metastable rigid solid to the stable unstressed solid at finite and sufficiently small strain rates may be described by a {\em nucleation} process. A parameter-free prediction of the strain rate dependent, mean, limiting deformation beyond which this nucleation occurs and a rigid crystal first begins to flow, is one of the verifiable outcomes of our work.  

Essential for these findings is the identification of a thermodynamic variable $X$, the order parameter of the transition, and its conjugated field $h_X$, which we define shortly.  We show that thermodynamically stable rigid solids exist for finite, negative $h_X$. In addition, we also obtain a line of first order transitions from a rigid solid to a solid state with zero stress. In the thermodynamic limit, this phase boundary extrapolates to $h_X\to 0^-$ giving rise to the aforementioned, experimentally observable transition associated with stress relaxation. We thus follow here a procedure analogous to many other condensed matter systems (for a classic example see~\cite{griffiths}) wherein deeper insight is obtained, leading to quantitative predictions, by first introducing a field $h_X$ and then letting $h_X \to 0$ after taking the thermodynamic limit.

\begin{figure}[t]
\begin{center}
\includegraphics[width=0.35\textwidth]{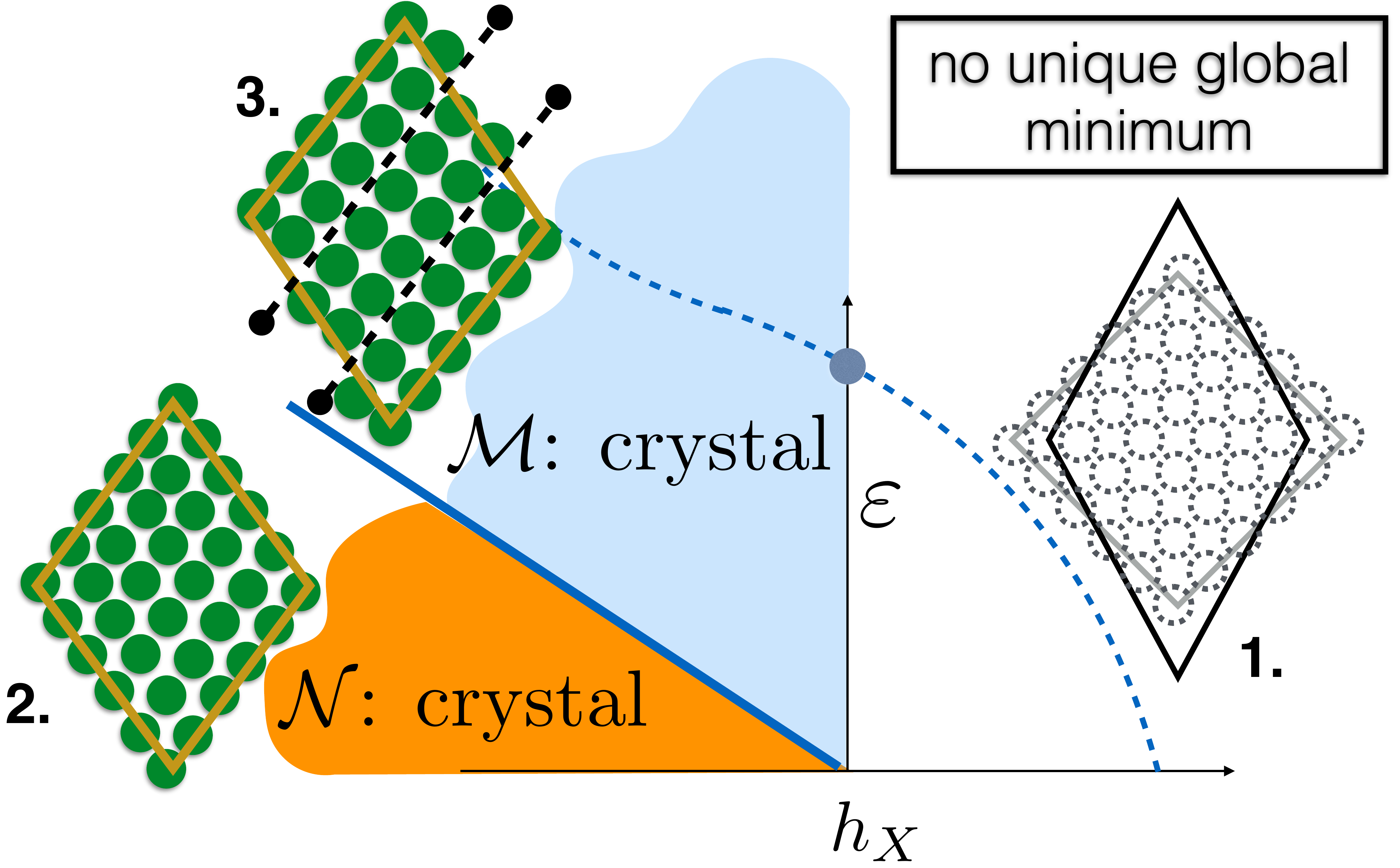}
\caption{\label{schema} Schematic phase diagram in the $h_X$ - pure shear, $\varepsilon$, plane showing regions of stability of the two (initially square) crystalline phases ${\mathcal N}$ (orange region) and ${\mathcal M}$ (blue region). Insets show $\varepsilon$ acting on a square (dashed line) box (1) and depict the ${\mathcal N}$ (2) and ${\mathcal M}$ (3) phases schematically; black dashed lines with end dots denote slip planes. The equilibrium phase boundary is shown as a blue solid line and the locus of the ${\mathcal N}\to{\mathcal M}$ dynamical transition as a blue dashed line. The intersection of the dashed line with the $h_X = 0$ axis is the conventional, strain rate dependent, yield point (grey filled circle).}
\end{center}
%\vskip -.5 cm
\end{figure}

Consider, therefore, a (single phase) crystalline solid completely enclosed by a deformable boundary. The solid is composed of macroscopic, classical particles, e.g. a colloidal crystal~\cite{colloid}. Our main conclusions are summarised in Fig.~\ref{schema} where we plot a schematic phase diagram of the crystalline solid under changes of boundary shape, parametrised by a pure shear or uniaxial strain ${\varepsilon}$, and $h_X$.  There are two distinct ways in which the solid may respond to $\varepsilon$, either resisting it by producing internal stress (the rigid ``normal'' ${\mathcal N}$ phase) or deforming plastically to conform to the shape of the boundary, expelling stress from the bulk (the ``Meissner'' ${\mathcal M}$ phase~\cite{tinkham, toledano}). Since the ${\mathcal M}$ phase deforms by slipping over an integral number of lattice spacings, crystallinity is preserved. The resulting ${\mathcal M}$ solid is {\em structurally identical} to the undeformed ${\mathcal N}$ crystal, save for the presence of surface steps. Each slip line, however, leaves in its wake a set of particles whose neighbourhoods have been {\em rearranged}. The field, $h_X$, assigns a bulk free energy cost for these rearrangements in a manner we describe below, and explicitly breaks the discrete, permutation symmetry, causing a first order transition and ${\mathcal N}$---${\mathcal M}$ phase coexistence. The phases co-exist across a phase boundary which extrapolates to $\varepsilon = 0$ as $h_X \to 0^-$. The ${\mathcal N}$ phase is metastable for all $\varepsilon$ on the $h_X = 0$ line and eventually decays by a nucleation process~\cite{sausset} with an $\varepsilon$-dependent rate.  For fixed observation time, this decay process manifests itself as a sudden drop of stress at some $\varepsilon = \varepsilon^*$, where plasticity initiates~\cite{rob}. This dynamical transition point extends into a smooth transition line in the $h_X - \varepsilon$ plane that intersects the $h_X = 0$ axis at the observable value, $\varepsilon^*$. We show that thermodynamic parameters obtained from our equilibrium study may then be used to predict time-dependent, dynamic properties of this transition.

%The rest of the paper is organised as follows. In the next section (Section~\ref{Tzero}) we define the collective coordinate $X$ and its conjugate field $h_X$ and then show how a simple energy audit leads to the $T=0$ phase diagram. This is followed (Section~\ref{equib}) by our results for the equilibrium phase transition at $T > 0$ showing coexistence of rigid and non-rigid solids at zero strain rate. Non-equilibrium deformation (Section~\ref{deformation}) of rigid solids at  finite strain rates is discussed next emphasising how these processes may be understood using quantities obtained from equilibrium studies. The paper ends with a discussion and conclusion (Section~\ref{dnc}) and a set of appendices where we give some details of our analytic derivations, simulations and the model studied. 
%
%
%\section{Collective coordinate, conjugate field and $T=0$ phase diagram}
%\label{Tzero}
The field couples to a collective coordinate $X = N^{-1}\sum_{i=1}^N \chi_i$ in the Hamiltonian $ {\cal H}  =  {\cal H}_{0} +  {\cal H}_{X} =  {\cal H}_{0} - N h_{X}X$, with ${\cal H}_{0}$ representing terms in the Hamiltonian that do not depend explicitly on $X$. For particle $i$, the local positive definite quantity $\chi_i$ with dimensions of length squared is the least squares error~\cite{falk} made by fitting a local affine deformation to relative displacements within a coarse graining volume surrounding particle $i$. The deformation is measured from a set of fixed reference coordinates. In a series of papers~\cite{sas1,sas2,sas3,sas4,sas5,sas6} (see also Appendix~\ref{NAP} for details) some of us have worked out in detail the statistical thermodynamics of $\chi_i$, which quantifies the ``non-affine'' component of the particle displacements, analytically and numerically for a number of two dimensional (2d) crystals at finite temperature. At any $T > 0$, $X$ behaves as a regular thermodynamic variable with a well-defined mean and variance $\sim N^{-1}$. Apart from this thermal contribution, $X$ also tracks local non-affine rearrangements of particles such as those resulting from the creation of defects~\cite{sas2,sas6}. The ensemble average $\langle X\rangle$ can be tailored using $h_X$ consistent with standard fluctuation response relations~\cite{sas2} and $h_X$ can also modify the probability of defects. Note that since $X$ is defined in terms of relative displacements, the term proportional to $h_X$ in ${\cal H}$ does not explicitly break translational invariance~\cite{popli}.

Positive values of $h_X$ help create non-affine rearrangements away from the reference configuration. Specific rearrangements, such as a slip by a lattice spacing, map the crystal onto itself and do not change lattice symmetry but still contribute to the energy ${\mathcal H}$ for non-zero $h_X$. Since $X$ has an upper bound $\sim L^2 \propto N^{2/d}$ where $L$ is a typical linear size, ${\mathcal H}$ and the corresponding free energy is unbounded below ($\sim-N^{1+2/d}$) in the thermodynamic limit. Therefore, there is {\em no well-defined global free energy minimum} for $h_X> 0$ although multiple local minima may exist as long-lived metastable states. At $h_X = 0$, of course, all states differing only in their value of $X$ are degenerate.  

On the other hand, a negative $h_X$ suppresses rearrangements and makes the reference configuration the thermodynamically stable phase at $\varepsilon = 0$. As $\varepsilon$ increases at constant $h_X$, there is a possibility of an equilibrium first order transition which may be understood from the following $T=0$ argument. In the ${\cal N}$ phase, $\varepsilon$ is the elastic strain~\cite{CL} and the bulk energy density is $\Delta E = \frac{1}{2} \sigma \epsilon = \frac{1}{2} G \epsilon^2$, where $\sigma$ is the elastic stress and $G$ is an elastic modulus. In the ${\cal M}$ crystal  non-affineness proportional to $|\varepsilon$, $X = \ell^2 |\varepsilon|$ with some lengthscale $\ell$, is produced instead by slipping of lattice planes and $\Delta E = -\rho h_X \ell^2 |\varepsilon|$, except for a surface contribution arising from steps that are formed as a consequence of the slips. Here, $\rho$ is the number density of the solid. Note that $X=0$ in the ${\cal N}$ solid~\cite{sas1} while $\sigma = 0$ in the ${\cal M}$ solid at coexistence. Equating, we get the ${\mathcal N}$---${\mathcal M}$ coexistence boundary as $ -h_X = G |\varepsilon|/ 2 \rho \ell^2$ (Fig.~\ref{schema}).  The ``thermodynamic stress'' $\varsigma = \partial_{\varepsilon} \Delta E \neq 0$ for {\em both} ${\mathcal N}$ and ${\mathcal M}$ phases.

How do entropic contributions alter these arguments? Consider the case $\varepsilon = 0$ and $h_X < 0$. Permuting particles over a distance $R$ should give a configurational entropy gain scaling as $\sim N \log R$. The energy penalty for this, given that $X \sim R^2$, will be $\sim |h_X| N R^2$. Minimizing the free energy $-T N \log (R/a)+N |h_X| R^2$ gives $R \sim (T/|h_X|)^{1/2} $. Once $R$ falls below some fixed small value (say, the lattice constant $a$) this argument breaks down, implying that for $|h_X|> {\rm const} \times T$ entropic effects may be neglected and the $T=0$ considerations hold. For smaller $|h_X|$, however, rearrangements are possible up to a cutoff distance $\sim (T/|h_X|)^{1/2}$ that diverges as $h_X \to 0^-$. Once rearrangements are thermodynamically favoured, kinetic considerations become important. In particular, if rearrangements happen only by diffusion, then the associated {\em timescales} in solids are very large~\cite{rob} so that spontaneous transitions between different free energy minima, corresponding to distinct rearrangements, will become effectively unobservable within realistic times. We return to the question of dynamics later. 

Up to now, our discussion has been quite general and works for any crystal in any dimension. 
We now specialise to the case of the crystalline 2d LJ solid to study the ${\mathcal N}$---${\mathcal M}$ transition at $T > 0$ in detail. Accurate numerical results may be obtained for this case within reasonable computational times. Further, our results have experimental consequences. These may be relatively easily verified for 2d colloidal crystals~\cite{colloid}, for which a LJ interaction is a plausible model. 
\begin{figure}[t!]
\begin{center}
\includegraphics[width=0.48\textwidth]{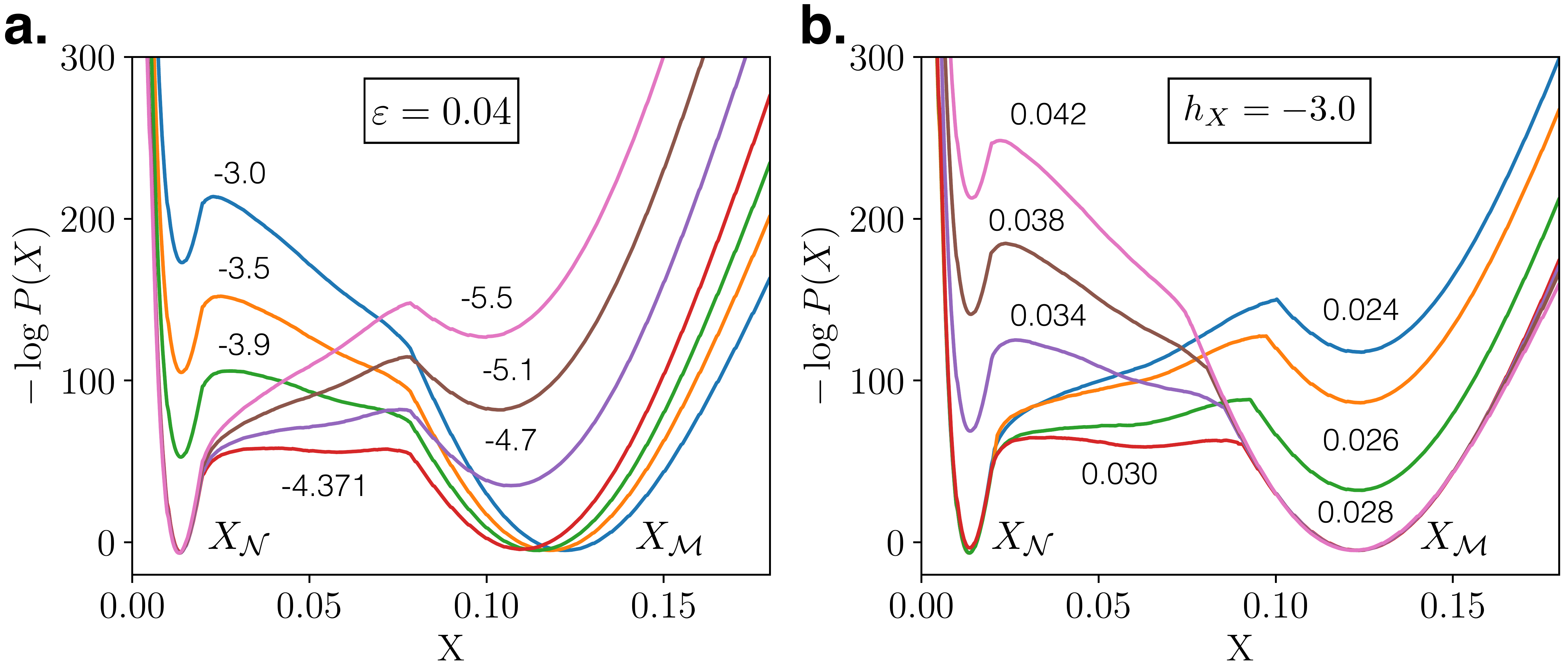}
\caption{\label{PX} Plots of the dimensionless free energy, $-\log P(X)$, at $T = 0.8$ and density $\rho=1.1547$ (i.e.\ lattice parameter $a=1.0$) as a function of $h_X$ at fixed $\varepsilon = 0.04$ ({\bf a}) and as a function of $\varepsilon$ at fixed $h_X = -3.0$ ({\bf b.}) as obtained from sequential umbrella sampling of 2d LJ crystals for $N = 1024$. The minima at small (large) $X$ values represent the ${\mathcal N}$ (${\mathcal M}$) phase. A first order ${\mathcal N} \to {\mathcal M}$ transition occurs as $-h_X$ or $\varepsilon$ is increased. The numbers on the graphs correspond to values of either $h_X$ ({\bf a})  or $\varepsilon$ ({\bf b}). Note the high barriers ($\sim 50 k_B T$) between the phases at coexistence.}
\end{center}
%\vskip -.7 cm
\end{figure}

\section{The equilibrium first order phase transition}
\label{equib}
We use a shifted and truncated LJ potential (see Appendix~\ref{LJ}) and standard LJ units for length, energy and time~\cite{frenkel}. For our results of the equilibrium structures and transitions at $T > 0$ we employed the sequential umbrella sampling (SUS) technique (see Appendix~\ref{SUS}) coupled to Monte Carlo~\cite{binder,SUS} in the constant number $N$, area $A = L_x \times L_y$, $\varepsilon$ and temperature $T$ ensemble.  Advanced sampling techniques such as SUS are necessary to overcome the large barriers between the ${\mathcal N}$ and ${\mathcal M}$ phases, enabling the equilibrium transition to be observed~\cite{sas4}. We show results for $T = 0.8$ and density $\rho=1.1547$, corresponding to the choice $a=1.0$ for the lattice parameter. Other $T$ and $\rho$ far from the 2d LJ melting line give similar results. Finally, our results for finite $N$ are extrapolated to draw conclusions on the equilibrium transition in the thermodynamic limit.

The main output of the SUS calculations is an accurate estimate of $P(X)$, the equilibrium probability distribution of $X$ as a function of $h_X$ and $\varepsilon$. In Fig.~\ref{PX}{\bf a} and {\bf b} we plot $-\log P(X)$, the free energy in units of $k_B T$ where $k_B$ is the Boltzmann constant. To obtain these results we use the efficient histogram reweighting method~\cite{borgs90,ferrenberg88} starting from a few chosen $h_X$ and $\varepsilon$. The two minima at $X_{\mathcal N}$ and $X_{\mathcal M}$ correspond to the two competing phases, with a first order transition from ${\mathcal N}$ to ${\mathcal M}$ occurring as a function of either $h_X$ or $\varepsilon$. The barrier between the phases at coexistence is high and hence the phase transition is impossible to observe using standard simulation techniques. The large value of $X_{\mathcal M}$ results from a finite density of percolating slip bands with large local $\chi$. Configurations for $X_{\mathcal N} < X < X_{\mathcal M}$ at coexistence show mixed phases similar to other systems with first order transitions~\cite{tfot}.  
\begin{figure}[t]
\begin{center}
\includegraphics[width=0.49\textwidth]{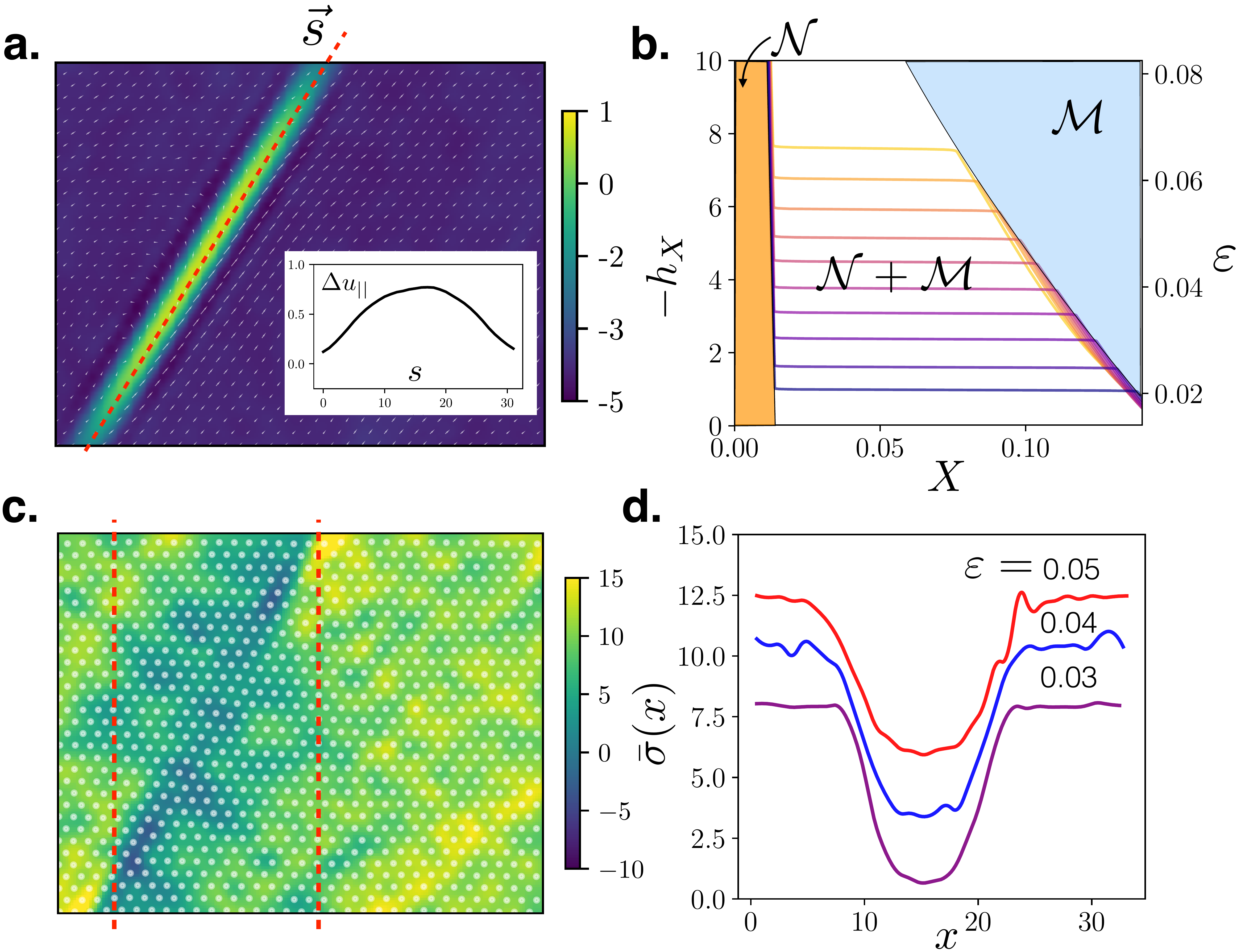}
\caption{\label{movie} {\bf a.} Plot of the local $\log (\chi)$ values (colour map) superimposed on the particle displacements (grey arrows) for a $N = 1024$ 2d LJ solid at $\varepsilon = 0.05$ and $h_X^{\rm coex} = -5.07$. The data was averaged over $500$ configurations corresponding to $X = 0.0358$ where a mixed phase configuration is observed. Note the presence of a slip band (red dashed line) with large $\chi$. Inset shows $\delta u_{\parallel}$, the relative displacement of particles along the direction of the slip band, consistent with that expected from a pair of dislocations with opposite signs and large overlapping cores. {\bf b.} Equations of state $-h_X$ (left axis) vs $X$. These are shown for several fixed $\varepsilon$ but overlap to a large extent, except where jumps in $X$ produce horizontal tielines (labelled by the corresponding $\varepsilon$ on the right). The large jump between the coexisting phases indicates a strongly first order transition. {\bf c.} Plot of the local stress $\sigma$ (colour map) superimposed on the particle positions (grey circles) for the same set of configurations as in ({\bf a}) showing a prominent stress interface (red dashed lines). {\bf d.} Plot of the stress $\bar{\sigma}(x)$ averaged in the vertical $y$ direction for three values of $\varepsilon$ along the phase boundary. The interface between the ${\mathcal N}$ phase (high stress) and the ${\mathcal M}$ phase (low stress) is clearly visible.}
\end{center}
\vskip -.2 cm
\end{figure} 

In Fig.~\ref{movie}{\bf a} we show a mixed phase configuration at coexistence where a portion of the solid slips locally, decreasing stress. This is apparent from the map of local $\chi$ values, which are largest at the slip band. The slip band is composed of a ``proto'' dislocation dipole with a large overlap between the individual defect cores~\cite{rob} lying on one of the close packed atomic lines of the triangular lattice. Scanning over configurations for $X$ between $X_{\mathcal N}$ and $X_{\mathcal M}$ reveals a slip band of increasing linear size until it percolates the whole solid, wraps about the periodic boundaries a few times, commensurate with the aspect ratio of the box and finally annihilates with itself at $X \approx X_{\mathcal M}$. In a periodically repeated scheme, therefore, this configuration corresponds to a finite slip band density $\sim {\mathcal O}(1)$. The proportion of the two phases follow a lever rule typical of first order transitions~\cite{tfot}. The strongly first order nature of the transition is obvious from the equations of state $X$ vs $h_X$ at fixed $\varepsilon$ obtained by plotting the expectation value of $X$ computed from the $P(X)$ in Fig.~\ref{movie}{\bf b}. The phase transition for each $\varepsilon$ is shown by a horizontal tieline, and labelled by $\varepsilon$ on the right axis. The end points of the tielines thus also give the $X$-$\varepsilon$ phase diagram. While the discontinuity in $X$ at the transition appears to decrease with increasing $\varepsilon$, it cannot vanish since a slip band always creates local $\chi$. 
\begin{figure*}[ht!]
\begin{center}
\includegraphics[width=0.95\textwidth]{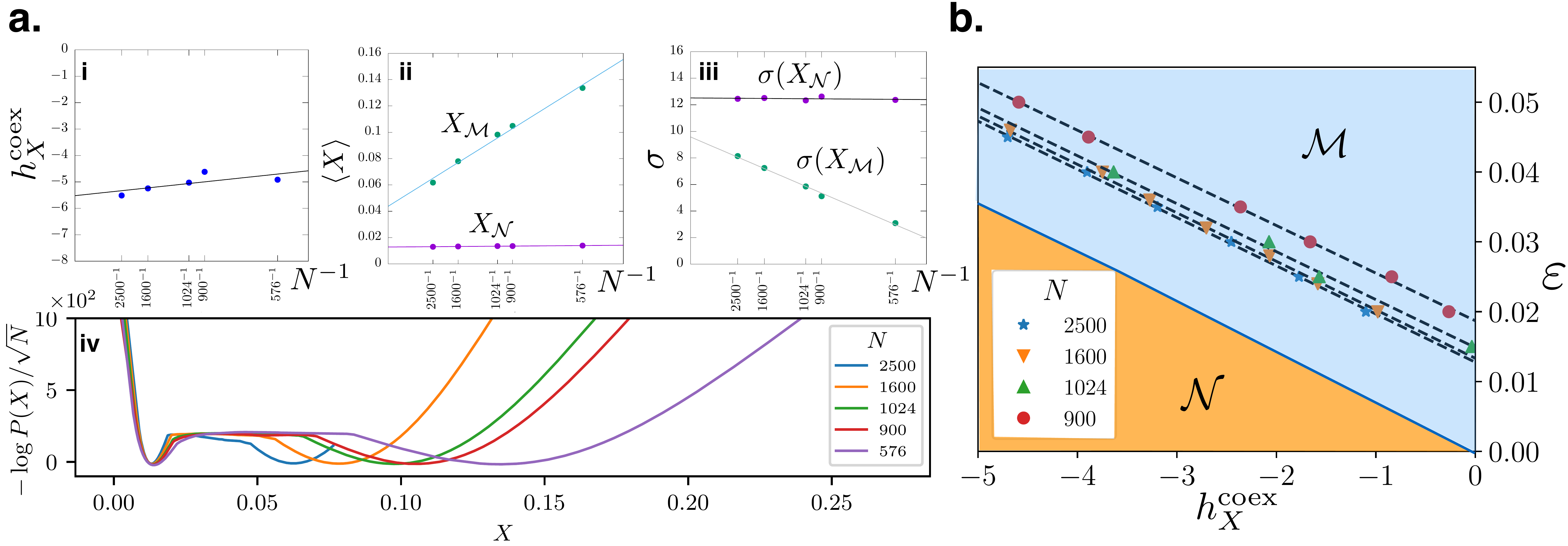}
\caption{\label{FSS} {\bf a.}~Finite size scaling at $\varepsilon=.05$ of the value of $h_X^{\rm coex}$ ({\bf i}), the coexisting $X_{\mathcal N}, X_{\mathcal M}$ at the two minima in $-\log P(X)$ ({\bf ii}), the coexisting stress values across the ${\mathcal N} - {\mathcal M}$ interface ({\bf iii}) and collapse of the scaled surface free energy in $-\log P(X)$ ({\bf iv}).~{\bf b.} The $h_X - \varepsilon$ phase boundary showing that as $N\to \infty$, the boundary tends to intersect the origin (schematic: orange and blue shaded regions and blue line as in Fig.~\ref{schema}).}
\end{center}
%\vskip -.5 cm
\end{figure*}

The nature of the ${\mathcal M}$ phase for a finite sized box is complex. While a deformation $\varepsilon$ needs a linear density of slip bands $\sim |\varepsilon|/a$ for complete stress relaxation, the number of slip bands actually observed depends on the size as well as the shape of the box. For example, if the required linear density of slip bands is below $L^{-1}$ then no bands can be accommodated and stress cannot be relaxed, i.e.\ the ${\mathcal M}$ phase cannot exist even if it is the stable phase in the thermodynamic limit. 
In general, there may be multiple minima in $X$ associated with distinct {\em families} of parallel slip bands at crystallographically allowed angles. At large $\varepsilon$, these further transitions involving additional slipping occur at values of $X$ higher than those shown here. Phases with a larger density of slip lines have lower $\sigma$, down to $\sigma = 0$. The exact sequence of these higher order transitions depends on the details of the simulation box. We do not pursue this here as the deformation \emph{dynamics} and departure from rigidity will be determined by the kinetics of the first transition.

In Fig.~\ref{movie}{\bf c} we return to the mixed phase configuration shown in {\bf a} and study the local internal stress $\sigma$ conjugate to $\varepsilon$ superimposed on the particle positions. A prominent {\it interface} between the two coexisting phases is clearly seen. The ${\mathcal M}$ phase eliminates stress from its bulk by particle rearrangements, i.e.\ slip, as expected, while stress is retained in the bulk of the ${\mathcal N}$ phase. As the amount of the second phase grows, total stress is proportionately reduced.

Capillary fluctuations of this interface~\cite{tfot} around the mean position are also seen in Fig.~\ref{movie}{\bf c}. Averaging the local stress $\bar{\sigma}(x)$ in the vertical direction and plotting it as a function of the horizontal coordinate $x$ reveals an interface where these capillary fluctuations are averaged out. This is shown in Fig.~\ref{movie}{\bf d}. We have plotted $\bar{\sigma}(x)$ for a few values of $\varepsilon$ on the phase boundary. While the jump in $\sigma$ decreases with $\varepsilon$, the interface remains, nevertheless, sharp.  

We must emphasise here that this is an {\em equilibrium} interface between two {\em co-existing} phases with different values of $\sigma$ but both with bulk crystalline order. Such stable interfaces do not form at $h_X = 0$ for $\varepsilon \neq 0$ in the thermodynamic limit and have, therefore, never been described before. 

\subsection{Finite size scaling}~We carry out a finite size scaling~\cite{binder,tfot} analysis at $\varepsilon = 0.05$ to establish that the transition between the two phases is indeed first order. At a first order transition, corrections to order parameters and to the transition point scale as $\sim L^{-d}\sim 1/N$~\cite{tfot}. This is apparent from Fig.~\ref{FSS}{\bf a}~{\bf i}-{\bf iii} where the coexisting $h_X^{\rm coex}$, the values of $\langle X \rangle$ and stress $\sigma$ for the two phases show the expected scaling behaviour. Finite size corrections to the properties of the ${\mathcal M}$ phase are observed to be quite substantial due to the commensurability issues discussed before. To show that the free energy cost of creating two parallel interfaces (lines) scales as $L^{d-1} = \sqrt{N}$, we plot $-\ln P(X)$ obtained at coexistence for different $N$ using scaled coordinates (Fig.~\ref{FSS}~{\bf a}~{\bf iv}). The region corresponding to mixed ${\mathcal N}-{\mathcal M}$ configurations collapses onto a single horizontal line as expected. Finally the $T > 0$ equilibrium phase boundary in $h_X$ and $\varepsilon$ is shown in Fig.~\ref{FSS}{\bf b.} for different $N$. The phase boundary is quite linear showing that our rather simplistic $T=0$ calculation gives a qualitatively correct result. The offset in $\varepsilon$ at $h_X \to 0^-$ ($\sim \varepsilon$ produced by a single slip band) is expected to vanish in the thermodynamic limit. For a thermodynamically large solid, we obtain phase coexistence only for $h_X \leq 0$ with the phase boundary intersecting the origin i.e. $h_X = \varepsilon = 0$. \footnote{The phase diagram in the $h_X$ and $\varsigma$ plane can also be drawn using our SUS data but does not contain substantial new information.} 

This completes our description of the {\em equilibrium} first order, ${\mathcal N}$ --- ${\mathcal M}$ phase transition. The transition is {\em reversible} with $\varepsilon$, a thermodynamic variable, being applied quasi-statically, i.e.\ $\dot \varepsilon = 0$. We study below the implications of the equilibrium transition on the dynamics of deformation ($\dot \varepsilon \neq 0$).  

\section{Nucleation dynamics and plastic deformation}
\label{deformation}
In the limit of $h_X \to 0^-$, a macroscopically large rigid (${\cal N}$) solid is metastable for all $\varepsilon > 0$. This is illustrated in Fig.~\ref{barr}{\bf a} for a $N = 1024$ solid where we plot $-\log P(X)$ as a function of $\varepsilon$ extrapolated to $h_X = 0$. As $\varepsilon$ increases, therefore, the ${\cal N}$ solid may decay by a process in which nuclei of the ${\cal M}$ solid form (and grow) within the body of the ${\cal N}$ phase. At the end of this process, the equilibrium, stress free,  ${\cal M}$ crystal thus formed is {\em identical} in all respects to the unstrained ${\cal N}$ crystal and with its other infinitely many copies differing only by their values of $X$. We show below how the free energies calculated using SUS may be used to study the dynamics of this nucleation process in quantitative detail. 
 
\subsection{Nucleation barriers}
\begin{figure}[t]
\begin{center}
\includegraphics[width=0.49\textwidth]{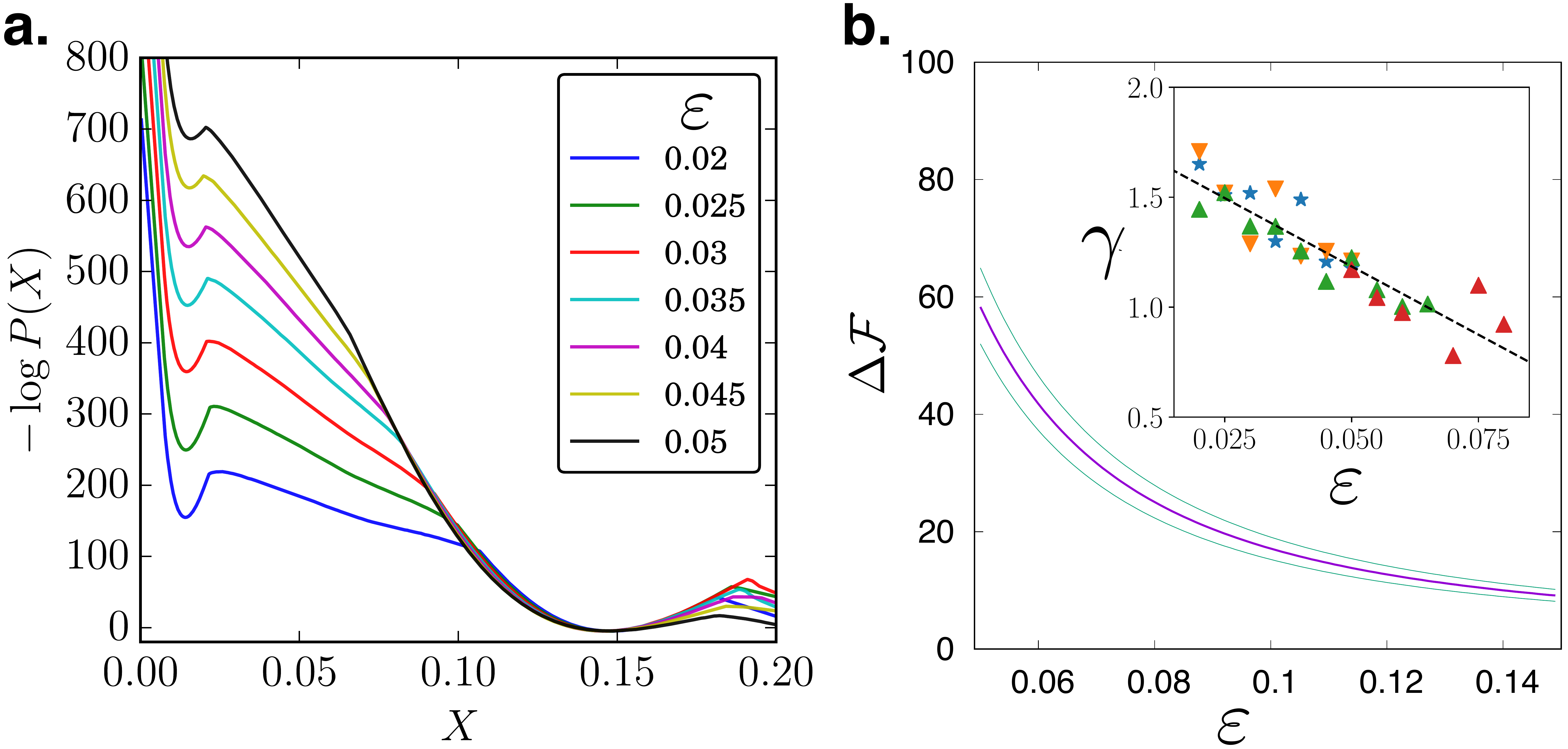}
\caption{\label{barr} {\bf a.}~Equilibrium free energy from SUS for a $N=1024$ solid for various $\varepsilon$ at $h_X = 0$ for comparison. {\bf b.}~The nucleation barrier $\Delta{\cal F}$ (solid purple line) as a function of $\varepsilon$, the cyan lines mark the uncertainties of our results. Inset shows the interfacial free energy $\gamma$ for various $N$ obtained from SUS (symbols have same meaning as in Fig.~\ref{FSS}{\bf b})  plotted against $\varepsilon$ together with the linear fit $\gamma = -12.420 \varepsilon + 1.806$.}
\end{center}
%\vskip -.5 cm
\end{figure}

~Following standard classical nucleation theory~\cite{CL, inter, sausset, becker} (CNT)  we write the dimensionless excess free energy of a configuration containing a circular droplet of the ${\cal M}$ phase of size $R$ surrounded by the stressed ${\cal N}$ crystal as $\Delta {\cal F} = -\pi R^2 \Delta_b  + 2 \pi R \gamma$. The first term represents the bulk free energy gain, which assuming that the ${\cal M}$ solid is stress free, gives $\Delta_b = \frac{1}{2} \sigma \varepsilon/(k_B T)$. The second term involves the equilibrium interfacial free energy $\gamma$ between ${\cal N}$ and ${\cal M}$ phases. Since equilibrium interfaces exist only at coexistence~\cite{tfot, inter}, we need to obtain $\gamma$ from $-\log P(X)$ along the phase boundary. The height of the horizontal region in $-\log P(X)$ (see Fig~\ref{FSS}~{\bf a}~{\bf iv}) relative to the depth of the minima is given by $2\gamma L_y + $ subdominant ($\sim {\mathcal O}(\log L)$ etc.) contributions.  Factoring out the length of the pair of parallel interfaces then gives the {\em finite size scaled} value for $\gamma$, which is shown in the inset of Fig.~\ref{barr}{\bf b} as a function of $\varepsilon$ for various $N$. Since the subdominant contributions are small and unobservable, we use a linear fit through the data to obtain $\gamma(\varepsilon)$ along the phase boundary. Extrapolation to $\varepsilon = 0$ gives $\gamma = 1.8 \pm 0.1$ as the surface free energy of the infinite solid, which enters the CNT calculation for nucleation of the ${\cal M}$ phase at $h_X = 0$. The size of the critical nucleus is $R_c = \gamma/\Delta_b$ and the nucleation barrier is $\Delta {\cal F} = \pi \gamma^2/\Delta_b$. The latter  is plotted as a function of $\varepsilon$ in Fig.~\ref{barr}{\bf b} as a solid curve. The free energy barrier has been approximately modelled earlier using specific, correlated defect structures such as arrays of dislocation loops~\cite{sausset, bruinsma, yip}, at $h_X = 0$. The full dynamical problem of many interacting dislocations in crystals is complex and remains unsolved~\cite{rob,yip}, necessitating many simplifying assumptions. In our description $\gamma$ is obtained without assuming any specific dislocation structure. 

We may now obtain the {\em mean first passage} nucleation time as $\tau_{FP} = \tau_0 \exp (\Delta {\cal F})$, in the limit of large $\Delta {\cal F}$ where $\tau_0$ is a relevant time scale~\cite{CL, becker}. Formally, $\tau_0$ is the time taken for nucleation when the barrier vanishes, but this interpretation is problematic because in that limit the nucleation picture itself fails.  We show later how this $\tau_0$ may be extracted from molecular dynamics (MD) data. Since the ${\mathcal N}$ phase is always metastable, $\tau_{FP}$ is finite for all $\varepsilon > 0$ and diverges as $\varepsilon \to 0$. Thus, if one waits long enough, a transition from ${\mathcal N} \to {\mathcal M}$ is inevitable at any $\varepsilon \neq 0$. We compare the CNT estimate to measured nucleation times in MD simulations below.  

\subsection{Molecular dynamics}~We perform MD simulations~\cite{allen, frenkel} (see also Appendix~\ref{MD}) for $128\times128 = 16384$ LJ particles in the $NA$(shape)$T$ ensemble at the same density as the SUS calculations for $0.2 < T < 1.2$.  The direct way of comparing SUS and MD is to compute transition times by holding the solid at various strain values. This protocol has technical issues because applying a finite strain suddenly to a solid causes transient shock waves that make extraction of meaningful data impossible. In the molecular dynamics simulations, therefore, the strain is ramped up from zero 
 in steps of $\Delta \varepsilon$, waiting for a time $t_W$ to obtain an average strain rate $\dot \varepsilon = \Delta \varepsilon /t_W$. We look for a drop in stress to mark the beginning of plasticity at the yield point $\varepsilon^*$.

\subsection{$X$ as a reaction coordinate}~To proceed any further, we must first establish that the plastic event at $\varepsilon^*$ does indeed represent an ${\mathcal N} \to {\mathcal M}$ transition described by $X$. In other words we need to show that $X$ is also the relevant {\em reaction coordinate} for deformation.

\begin{figure}[t]
\begin{center}
\includegraphics[width=0.49\textwidth]{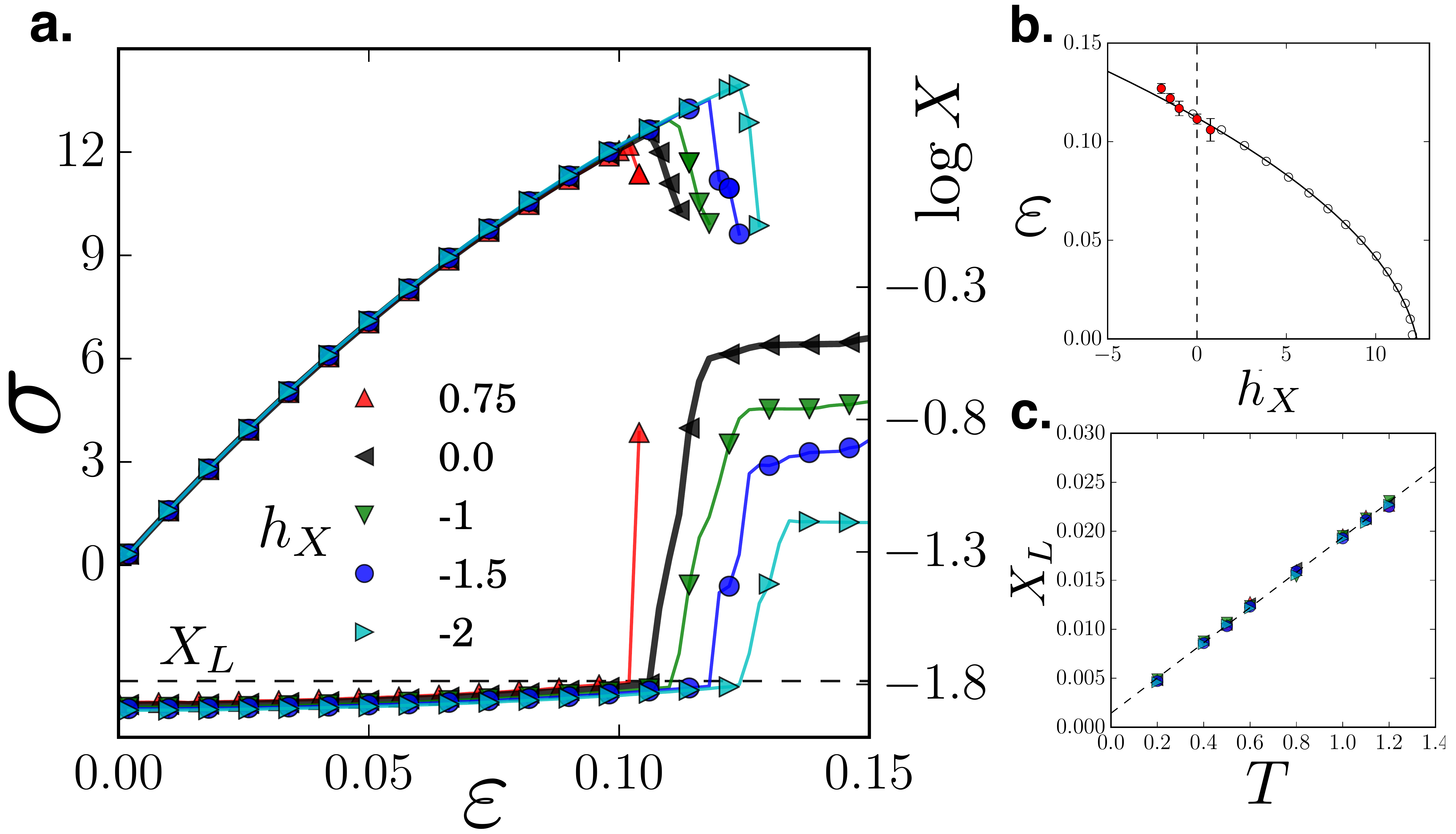}
\caption{\label{yield} {\bf a.}~Plots of $\sigma$ (axis on left) and $\log \langle X \rangle$ (axis on right) as a function of $\varepsilon$  from MD simulations at $\dot \varepsilon = 3.33\times 10^{-5}$. Note that the jumps in $\sigma$ and $\langle X \rangle$ at the dynamical transition coincide. The dashed line shows the limiting value $X_L = 0.0165$ before the transition.  {\bf b.} Plot of the ${\mathcal N} - {\mathcal M}$ dynamic transition in the $(\epsilon,h_{X})$ plane. The open symbols mark the dynamical transition $\epsilon^{*} (h_X)$ predicted from LRT calculations assuming $\langle X \rangle \to X_L$. The black curve is a parabola fitted through the points. The red filled circles are transition points obtained from MD simulations.  {\bf c.} Scaling of $X_L$ with temperature showing that $X_L /T $ is a constant.  }
\end{center}
%\vskip -.5 cm
\end{figure} 

The $\sigma(\varepsilon)$ curves obtained for various $h_X$, including $h_X = 0$, at a fixed value of $\dot \varepsilon$ are shown in Fig.~\ref{yield}{\bf a}. These show a large stress drop at $\varepsilon = \varepsilon^*$ while $\langle X \rangle$ increases at the same value of strain indicating that particle rearrangements occur. The angular brackets here denote a time average as well as an average over several initial conditions. The jump in $\langle X \rangle$ with a simultaneous drop in stress indicates that these particle rearrangements, at the same time, relieve stress. We find that $\langle X \rangle$ always attains the same value $\langle X \rangle = X_L $ with $X_N < X_L < X_M$ just before the transition regardless of $h_X$. As the solid is strained, $\langle X \rangle$ increases; when it reaches $X_L$, enough thermal energy is available to the solid in order to cross the ${\mathcal N} \to{\mathcal M}$ barrier and the ${\mathcal M}$ phase begins to nucleate.  

The ${\cal N}$ phase decays if {\em either} $h_X$ {\em or} $\varepsilon$ is increased. If $X$ is the relevant coordinate for this transition, then this single quantity should describe yielding regardless of $h_X$ and $\varepsilon$. In other words, $\langle X (h_X, \varepsilon) \rangle = X_L$ should trace out a unique curve in $h_X - \varepsilon$ space, beyond which the ${\cal N}$ crystal decays. Since $\langle X \rangle$ grows linearly with $h_X$ to leading order and quadratically with $\varepsilon$ in the ${\cal N}$ phase~\cite{sas1},  this curve is a parabola. In the ${\cal N}$ phase, we use linear response theory~\cite{sas2} to obtain the $h_X$ values where $\langle X \rangle \to X_L$ for any given $\varepsilon$. To do this we record fluctuations of $X$ and spatial correlations of $\chi$ at $h_X = 0$ for a series of $\varepsilon$ values starting from zero. Linear response then gives $\langle X(h_{X},\varepsilon)\rangle  =  \langle X(0,\varepsilon)\rangle +h_{X}\langle[\Delta\chi(0,\varepsilon)]^{2}\rangle \Sigma_{\bf R} C_{\chi(0,\varepsilon)}({\bf R},0) = X_L$, where $C_{\chi(0,\varepsilon)}({\bf R},0)$ is a two point correlation function. The predicted values for the location of the dynamical ${\cal N} \to {\cal M}$ transition, which follow the expected parabolic relation, are given as open symbols in Fig~\ref{yield}{\bf b}. Note that we now predict the location of a {\em plastic} event based on {\em equilibrium} thermal fluctuations of a thermodynamic variable in configurations corresponding to small $\varepsilon$ (and $h_X = 0$) where dislocations or other defects {\em may not even be present}. The filled symbols in Fig.~\ref{yield}{\bf b} show $\epsilon^{*} (h_{X})$ obtained from MD. The extrapolation of the calculated curve to $h_X = 0$ agrees extremely well with the measured value. At large negative $h_X$ the linear response prediction ceases to be valid. 
Given that in the ${\cal N}$-phase thermal fluctuations are primarily responsible for making $X$ non-zero~\cite{sas1}, $X_L$ should scale with temperature. Fig.~\ref{yield}{\bf c} shows that on varying only $T$, this expectation is justified and $X_L(T)/T$ is a constant.  
We expect this behaviour to be generic and easily verifiable in experiments on colloidal solids~\cite{colloid} constituting a stringent test for our theory.
\begin{figure}[t!]
\begin{center}
\includegraphics[width=0.49\textwidth]{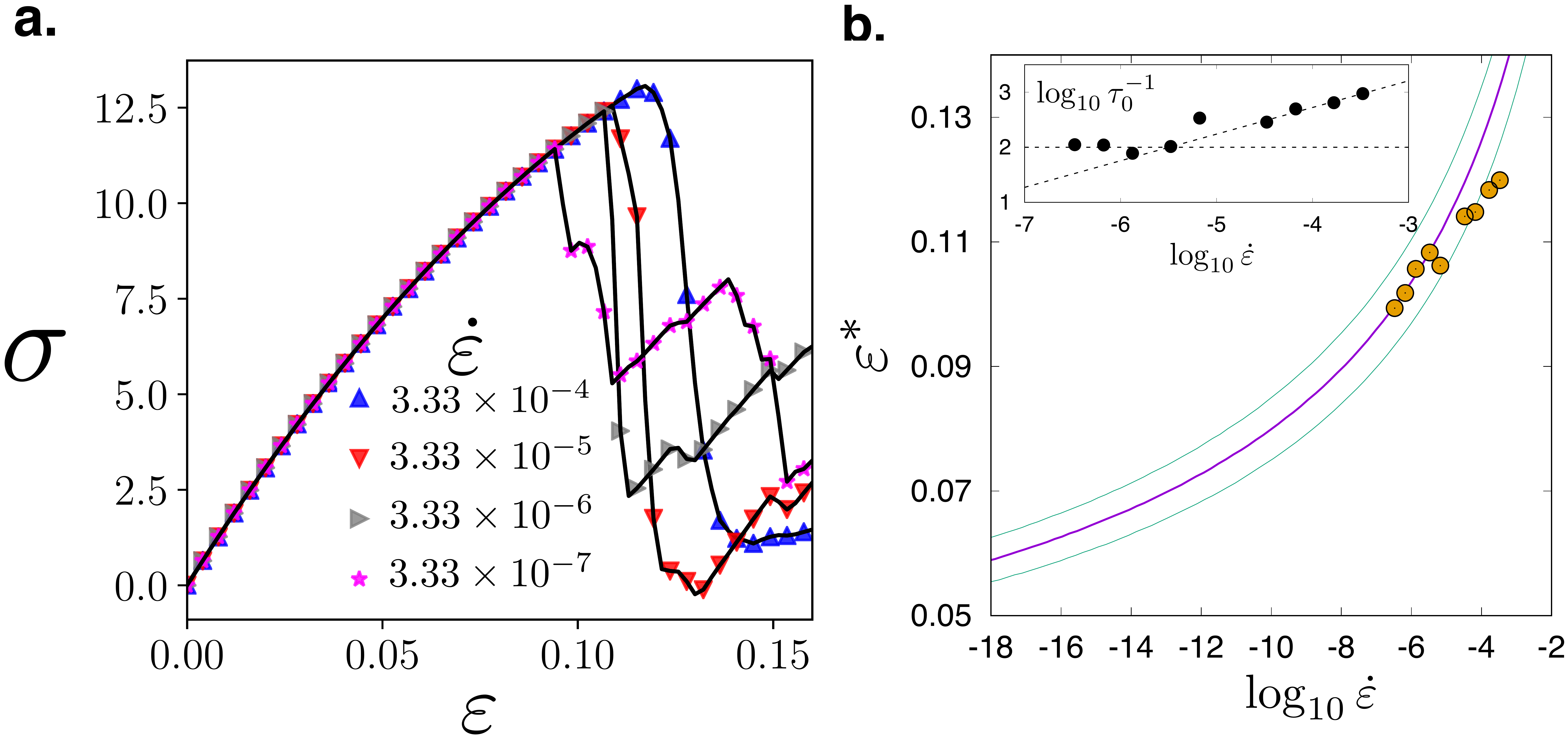}
\caption{\label{cnt} {\bf a.} Stress - strain curves from MD at $T = 0.8$ at $h_X = 0$ for four $\dot \varepsilon$ varying over three decades.
{\bf b.} The calculated value for the yield strain $\varepsilon^*$ as a function of the strain rate $\dot \varepsilon$ in LJ units. The lines have the same meaning as in Fig.~\ref{barr}{\bf b}. The points are from our MD results. Inset shows a plot of $\log_{10}\tau_0^{-1} = \log_{10}(\dot \varepsilon/\tilde{\dot \varepsilon})$ as a function of $\dot \varepsilon$ obtained from our MD. We have used a value of $\log_{10} \tau_0^{-1} = 2.0$ as obtained from the plateau value in the inset to compare our MD data with the SUS predictions.}
\end{center}
\end{figure}

Since $\varepsilon$ is ramped up from zero at a fixed rate, $\dot \varepsilon$, the dynamical transition i.e. the values of $\varepsilon^*$ and $X_L$, depend on $\dot \varepsilon$. We show now that this rate dependence may be predicted using parameters extracted from the equilibrium phase transition.

\subsection{Self consistent classical nucleation theory}~We are now in a position to compare predictions of CNT with MD results. In Fig~\ref{cnt}{\bf a}, we plot $\sigma(\varepsilon)$ curves at $h_X = 0$ for four deformation protocols where $\dot  \varepsilon$ varies over three decades. The yield point $\varepsilon^*$ is a function of  $\dot  \varepsilon$ and appears to vanish as $\dot  \varepsilon \to 0$. This result is consistent with the ${\mathcal N}$ phase being metastable for all $\varepsilon > 0$.

To obtain predictions for $\epsilon^*$ using the barriers obtained from our SUS calculations, however, we need a small modification of the theory. For the protocol followed in MD, the barrier is not constant, but varies as the strain is ramped up in time. If the variation is smooth and slow, we may use the self consistent formula $\tau_{FP} = \tau_0 \exp[\Delta {\mathcal F}(\dot \varepsilon \tau_{FP})]$~\cite{sc-cnt}. Noting that $\dot \varepsilon \tau_{FP} = \varepsilon^*$, we get a self consistency equation for $\varepsilon^*$ which needs to be solved numerically, $$ \Delta {\mathcal F}(\varepsilon^*) = \log \varepsilon^* - \log(\tilde{\dot \varepsilon}),$$ where $\tilde{\dot \varepsilon} = \tau_0 \dot \varepsilon$.  This  approximation should be valid in the small $\dot \varepsilon$ regime. Since small $\dot \varepsilon$ gives small $\varepsilon^*$, and therefore large barriers, the regimes of validity of the self consistent approximation and CNT itself coincide.   

In principle $\tau_0$ could be taken as a fitting parameter which links energies to time scales. But fitting $\tau_0$ to MD data is uncertain because the range of validity of CNT is not known \emph{a priori}. It has also been interpreted as the time taken for a dislocation dipole (in 2d) or loop (3d) to form~\cite{sausset}, although obtaining an estimate for $\tau_0$ using this interpretation requires additional assumptions. The nucleation rate of dislocations is a technologically important quantity and has been measured using experiments and computer simulations~\cite{schall,julie,julie2,dis-nuc}. For a real 3d solid, the pre-factor can be written as $\tau_0^{-1} = \Gamma f_c^{+}$ where $\Gamma$ is the Zeldovich factor and $f_c^{+}$ the ``attachment'' rate~\cite{becker}. These have been estimated for Cu single crystals and give $\tau_0 = 3.43\times10^{-14} {\rm s}$~\cite{dis-nuc}. This number is also consistent with $\tau_0^{-1}$ compared with typical Debye frequencies $\omega_D \sim 10^{13}~ {\rm Hz}$~\cite{CL}. In LJ units, where the unit of time is about a pico-second ($10^{-12}$), we obtain $\tau_0 = 0.0343$. Below we show how $\tau_0$ can be alternatively obtained by appealing to the internal consistency of CNT without using dislocation nucleation times as input. Remarkably, we also determine, at the same time, the range of validity of CNT for our MD data. 

From our self consistent CNT theory we obtain $\tilde {\dot \varepsilon}$ for each $\varepsilon^*$ obtained from MD. We now use the fact that $\tau_0$ should be independent of $\dot \varepsilon$ if CNT is valid and obtain $\tilde {\dot \varepsilon}/\dot \varepsilon$ for each $\varepsilon^*$. This is shown in the inset of Fig.~\ref{cnt}{\bf b}. Note that for small $\dot \varepsilon$ we obtain a plateau in the values of $\tau_0 = 0.01$ thus calculated. Deviations from the plateau value begin from $\dot \varepsilon \approx 10^{-5}$ or $\varepsilon^* \approx 0.11$. Comparing with Fig.\ref{barr}{\bf b} we observe that this corresponds to a barrier height of about $10-20 ~{\rm k_B T}$, which is completely consistent with expectations. The data from MD is compared with the results of the self consistent theory, using $\tau_0$ obtained from the plateau value in the limit $\dot \varepsilon \to 0$, in Fig.~\ref{cnt}. Our SUS and MD data are in excellent agreement for the smallest $\dot \varepsilon$ values showing that the decay of the metastable ${\mathcal N}$ solid sets the time scale for microscopic processes responsible for stress relaxation. Our $\varepsilon^*(\dot \varepsilon)$ curve therefore is a prediction for the yield point at strain rates that are relevant for slow deformation of solids under experimental conditions~\cite{rob}. Such processes are impossible to probe in standard MD simulations and it is remarkable that SUS allows us access to these regimes. Although we have presented results for a single temperature $T = 0.8$, our predictions follow from the identification of $X$ as the reaction coordinate, with $X_L$ as its limiting value in the ${\mathcal N}$ phase. Since $X_L$ scales simply with $T$, we expect our $\varepsilon^*(\dot \varepsilon)$ to do the same as long as the temperature (and density) is not close to melting. Note that there are {\em no adjustable parameters} in our calculation.

\section{Discussion and conclusions}
\label{dnc}
We began our investigation by asking whether rigid solids can ever exist as a stable thermodynamic phase. We introduced a new collective variable $X$ that keeps track of non-affine particle rearrangements by comparing positions of distinguishable particles with a set of reference coordinates. By turning on a field $h_X$ conjugate to $X$ one can bias particle rearrangements and explicitly break permutation symmetry. We show that the breaking of this discrete symmetry leads to a first order transition and phase coexistence between a rigid solid and one where particles rearrange to eliminate stress. The first order transition is quite conventional in all respects and both bulk and surface properties of the coexisting phases scale in the expected manner. By measuring these properties in the presence of this fictitious field and subsequently taking both the thermodynamic limit and the limit $h_X \to 0$ we were able to obtain quantities that predict the dynamics of deformation of the solid in the absence of field without any fitting parameters. 

Irreversibility of plastic deformation is easy to understand. At large strain rates the product state is not in equilibrium and the process is irreversible. However, deforming a solid even with $\dot \varepsilon \to 0$ causes irreversibility if $h_X = 0$, due to the nature of the free energy landscape. Consider the reverse transformation from ${\cal M} \to {\cal N}$. Straining the ${\cal M}$ solid in the reverse direction now increases its free energy and it decays to a stable unstressed state. However all rearranged versions of the original ${\cal N}$ crystal have the same free energy and are equivalent candidate products. The solid at the end of the process is likely to reach one of these states instead of the original crystal, with overwhelmingly large probability, producing irreversible particle rearrangements. The large degeneracy of phases at $h_X = 0$ thus makes the reverse transformation non-unique and deformation irreversible. This is, of course, not true if $h_X < 0$ where equilibrium transformations between ${\cal N}$ and ${\cal M}$ are always reversible.   

So far, $h_X$ has been introduced as a device for understanding the relation between non-affine particle displacements and deformation in a solid with contact being made with experiments only at $h_X = 0$. For solids where individual particles can be distinguished and tracked, one should be able to realise $h_X$ in the laboratory and check our predictions in the full $h_X - \varepsilon$ plane. Indeed, it has already been discussed in detail~\cite{sas2, sas4, popli} how this may be accomplished in the future for colloidal particles in 2d using dynamic laser traps. Briefly, the set of reference coordinates is read in and a laser tweezer is used to exert additional forces, ${\cal F}_\chi ({\bf r}_i) = -\partial {\cal H}_X / \partial {\bf r}_i$ to each particle $i$ which bias displacement fluctuations. Since the additional forces depend on {\em instantaneous} particle positions, they need to be updated continuously through real time particle tracking. This is possible because timescales of colloidal diffusion are large~\cite{colloid}. We believe that this procedure will be achievable in the near future using current video microscopic and spatial light modulation technology~\cite{HOT}.  Colloidal ${\mathcal N}$ crystals stabilised under an artificially produced $h_X$ field should show new and interesting properties, such as high failure strengths with small elastic constants or vice versa, resistance to creep, very small defect concentrations etc. These properties, coupled to the fact that they are reversible and may be switched on or off or precisely tuned, may have some technological use.  

At large times, after the first plastic event at $\varepsilon^*$, a crystal under a constant deformation rate $\dot \varepsilon$ reaches a non-equilibrium steady state. Flow of a crystal with a vanishing strain rate $\dot \varepsilon \to 0$ may be understood as a succession of, perpetually occurring, ${\cal N} \to {\cal M}$ nucleation events, which cause deformation while attempting to reset stress to zero~\cite{sausset,rei}. In this regime, $\dot \varepsilon = \sigma/(G\,\tau_{FP})$, which may be interpreted as flow with a viscosity that diverges as $\sigma = G \varepsilon \to 0$.  At large strain rates, $\Delta {\cal F} \to 0$ so that nucleation ceases to be the relevant dynamical process. The ${\cal N}$ solid decays to a truly non-equilibrium steady state unrelated to ${\cal M}$.  Further, in this regime one obtains critical like behaviour and scale-free avalanche driven deformation~\cite{carmen, zaiser, scaling, sethna, jamming}. Our work suggests that such behaviour need not be universal but a consequence of large strain rates used in those studies.  

Our work is easily generalised to other realistic crystalline solids in 3d, for example Cu, Au or Al single crystals for which a wealth of experimental data already exists~\cite{dis-nuc,julie,julie2}. We can obtain $\gamma$ by turning on $h_X$ and subsequently set $h_X \to 0$ after taking the thermodynamic limit  to predict $\varepsilon^*(\dot \varepsilon)$ for these solids without adjustable parameters. The introduction of $h_X$, of course, does not depend on any particular model interaction used and can be implemented within any simulation scheme. It also does not change any of the elastic properties of these solids. Work along these lines is in progress and will be published elsewhere. 

In a departure from {\em all} known literature on plasticity of crystalline solids~\cite{sausset, rei, rob, yip, dis-nuc, carmen, zaiser, scaling, sethna, jamming, julie, julie2}, no property of dislocations such as core energies or dislocation - dislocation interactions~\cite{rob} etc. enter our discussion.\footnote{The equilibrium $\cal{N}$ and $\cal{M}$ phases are both defect free. For very large systems and at elevated temperatures a small defect concentration is expected on entropic considerations. Kinetic jamming effects may also contribute to increase the defect concentration in the $\cal{M}$ phase relative to $\cal{N}$.}   This fact imparts a greater range of applicability to our work than those explicitly based on the language of dislocations. We have used almost the same language to describe pleating of two dimensional sheets modelled as a network of permanently bonded vertices~\cite{sas4, debankur}. In such systems dislocations cannot form at all, although non-trivial fluctuations in the form of pleats can still be described using non-affine displacements. Similar phase coexistence between a stressed network and one where stress is relaxed by pleating is observed. 

A collective variable, similar to $X$ used by us, was
initially defined to characterize local particle rearrangements in glasses
under deformation~\cite{falk, falk-review,glassbook}. So our method, with perhaps
a few modifications and/or generalisations, could be also applied to
amorphous solids. The response of an amorphous solid to a deformation
in the zero strain rate limit ($\dot{\varepsilon}\to 0$) is, however,
expected to be very different from that of crystals. In this case, two
scenarios are possible: (i) For $\dot{\varepsilon}=0$, the system is not a
glass any more and behaves like a Newtonian fluid. This implies that no
broken symmetry is involved, such as the breaking of permutation
symmetry, associated with the flow of a crystal. So there is no
underlying first-order transition as in the crystal. (ii) The system is in an
ideal glass state. As a consequence, there is a nonzero yield
stress~\cite{chaudhuri13} $\sigma_{\rm y}$, i.e.~as a response to any
deformation with a given strain rate, the ideal glass state transforms
eventually to a flowing state with a finite stress $\sigma \geq \sigma_{\rm y}$
(at $\dot{\varepsilon}=0$, $\sigma=\sigma_{\rm y}$). The nature of
this transformation is an open issue (see, e.g., Ref.~\cite{barrat11}). Also
the onset of flow in a glass is different in nature from that of a crystal. While
in a crystal flow occurs via the formation of slip bands, in a glass flow is initiated by a percolating cluster of mobile regions~\cite{perc},
associated with a transition in the directed percolation universality class.
Furthermore, amorphous solids driven far from equilibrium under external
shear stresses exhibit complex deformation behaviour
\cite{PNAS2,manasa,denisov,itamar,chaudhuri13,barrat11,schall2,schall3,chaudhuri14,perc,shrivastav16,sentjabrskaja15}. 
A deeper understanding of all these issues is a challenge that we wish to pursue
using our methods in the future.

%\newpage

%\matmethods{
\appendix
%\section{Models, Formalism and Simulation Details} 
%\label{sec1}
%
%In this section, we commence our discussion by first introducing
%the non-affine field $h_X$ and the model Hamiltonians followed by a
%description of the simulation methodologies used.
%
\section{The projection formalism and the non-affine field}
\label{NAP}
Choose a reference configuration with $N$ particles where particle $i$ ($i=1,...,N$) has position ${\bf R}_i$. Displacing particle $i$ to ${\bf r}_i$ the
instantaneous position of the particle, produces ${\bf u}_i = {\bf r}_{i}-{\bf R}_{i}$. Within neighbourhood $\Omega$ around $i$, the relative displacements ${\bf \Delta}_{j} = {\bf u}_j-{\bf u}_i$ with respect to particle $j\neq i \in\Omega$. To obtain the ``best fit'' \cite{falk} local affine deformation ${\mathsf D}$ one minimises  $\sum_j [{\bf \Delta}_{j} - {\mathsf D}({\bf R}_{j}- {\bf R}_{i})]^2$ so that $\chi({\bf R}_i) > 0$ is the minimum value of this quantity. This procedure
also amounts to taking a projection~\cite{sas1} of ${\bf \Delta}_i$ onto a subspace defined by the projection operator ${\mathsf P}$ so that,
$\chi({\bf R}_i)= {\bf \Delta}^{\rm T}{\mathsf P}{\bf \Delta}$ where we use $\Delta$, the column vector constructed out of ${\bf \Delta}_i$. In ${\mathsf P} = {\mathsf I}-{\mathsf R}({\mathsf R}^{\rm T}{\mathsf R})^{-1}{\mathsf R}^{\rm T}$, the $Nd\times d^{2}$ elements of ${\mathsf R}_{j\alpha,\gamma\gamma^{\prime}} = \delta_{\alpha\gamma}R_{j\gamma^{\prime}}$ centering $\Omega$ at the origin. 
The global non-affine parameter, $X=N^{-1}\sum^N_i \chi({\bf R}_i)$ couples to $h_{X}$ in the Hamiltonian ${\cal H} =  {\cal H}_{0} - N h_X X$, with ${\cal H}_{0}$ as the Hamiltonian of any solid. Note that ${\bf u}_i \to {\bf u}_i + {\bf c}$, where ${\bf c}$ is an arbitrary translation, remains a symmetry of ${\cal H}$. The statistics of $\chi (h_X)$ and $X (h_X)$ may be computed using standard methods of statistical mechanics~\cite{sas1,sas2}.  

\section{The Lennard-Jones Model} 
\label{LJ}
The shifted and truncated LJ model is defined by $ {\cal H}_0 = \sum^N_{i=1} \frac{{\bf p}_i^2}{2 m} 
+ \sum_{i=1}^{N-1} \sum_{j>i} v_{\rm LJ}(r_{ij})$

with ${\bf p}_i$ the momentum and $m=1.0$ the mass of a particle.  The
interaction potential for a pair of particles, separated by a distance
$r$, is
$v_{\rm LJ} = 4\phi[(r_0/r)^{12} - (r_0/r_c)^{12} - (r_0/r)^6 + (r_0/r_c)^6]$
for $r \leqslant r_{c}=2.5r_{0}$ and $v_{\rm LJ}=0$ for $r>r_c$. Energy
and length scales of the LJ model are set by $\phi=1$ and $r_{0}=1$,
respectively. The unit of time is given by, $\tau = \sqrt{m r_0^2/\phi}$.

\section{Sequential Umbrella Sampling}
\label{SUS} 
SUS-MC~\cite{SUS,frenkel,binder} in the $NAT$ ensemble is implemented in a manner identical to that used in Ref.~\cite{sas4} using a periodically repeated rectangular box of dimensions $L_x \times L_y$. The range of $X$ is divided into small windows and sampled successively starting at $X=0$. We keep track of how often each value of $X$ within the $n$th window is realised and the resulting histograms $H(n)$ thus obtained are used to compute the probability $P(X)$. The SUS-MC runs were done for systems with $N = 2500, 1600, 1024, 900$ and $576$ LJ particles at $T = 0.8$ and the density $\rho=1.1547$ ($a=1.0$).  The entire range of $X$ (which varies depending on $N$)  is divided into $800-1000$ sampling windows with $\approx1\times 10^{8}$ MC moves attempted in each window. In each MC move, maximal displacements of $0.2~a-0.4~a$ along the $x$ and $y$ directions are allowed. 
The SUS-MC computations were done for various $h_X$ and $\varepsilon$. The simulation box is rescaled setting $L_x \to L_x (1 + \varepsilon)$ and $L_y \to L_y (1 - \varepsilon)$ which keeps the area constant upto linear order. The local stress is obtained from the SUS configurations in the usual way from averaging the virial, taking care of the three body terms implicit in the terms involving $h_X$~\cite{frenkel, sas4}.   

\section{Molecular Dynamics} 
\label{MD}
The MD simulations for the LJ were done for $128\times
128 = 16384$ particles within a periodically repeated box identical to that used in the SUS-MC at the same density $\rho=1.1547$ and several $T$ using a velocity Verlet algorithm~\cite{frenkel}. In most simulations, the MD time step $\delta t = 0.001$ in LJ time units; only near yielding, and for $h_X \ge 0.5$, a smaller time step of $\delta t = 0.0001$  is used. In the LJ case, the system is coupled to a Berendsen thermostat \cite{frenkel,allen}. The solid is first equilibrated in the absence of the $h_X$ for $t = 500$, followed by equilibration runs at different values of $h_X$, in each case for over  $t = 1000$. Pure shear is applied by rescaling the box in steps of $\Delta \varepsilon = 0.001$ with a waiting time $3 < t_W < 3000$ at each step. The mean strain rate $\dot \varepsilon = \Delta \varepsilon/t_W$. For each value of $h_X$ and $t_W$, $6-8$ independent runs were performed. 

%

%\end{materials}
%\showmatmethods

\acknowledgements
We thank M. Barma, M. Rao and S. Ramaswamy 
%and participants of the KITP summer school on ``The Physics of Glasses: Relating Metallic Glasses to Molecular, Polymeric and Oxide Glasses", 2010 
for discussions. 
%SS thanks the Okinawa Institute for Science and Technology for hospitality.
Funding from the FP7-PEOPLE-2013-IRSES grant no: 612707, DIONICOS is acknowledged.
%
%\showacknow
%%
%\newpage
%
%\pnasbreak

%


\begin{thebibliography}{10}

\bibitem{CL} 
Chaikin P and Lubensky T (1995) {\em Principles of Condensed Matter Physics} (Cambridge Press, Cambridge).
%
\bibitem{szamel1}
Szamel G and Ernst MH (1993) Slow modes in crystals: A method to study elastic constants \prb  48:112.
%
\bibitem{ruelle}
Ruelle D (1969) {\em Statistical Mechanics}, (Benjamin, New York).
%
\bibitem{penrose}
Penrose O  (2002) Statistical mechanics of nonlinear elasticity {\em Markov Processes Relat. Fields} 8:351.
\bibitem{sausset}
Sausset F, Biroli G and Kurchan J (2010) Do Solids Flow?  {\em J. Stat. Phys.} 140:718.
%
\bibitem{shibu}
Shaw S and Harrowell P (2016) Rigidity in Condensed Matter and Its Origin in Configurational Constraint \prl 116:137801.

\bibitem{rob}
Phillips R (2004) {\em Crystals, defects and microstructures: Modeling across scales} (Cambridge Press, Cambridge).

\bibitem{rei}
Barnes HA (1999) The yield stress - a review or `$\pi\alpha\nu\tau\alpha$~$\rho\epsilon\iota$' - everything flows? J. Non-Newtonian Fluid Mech., 81(1-2): 133-178.

\bibitem{griffiths} 
Griffiths R B  (1970) Thermodynamics near the two-fluid critical mixing point in He$^3$-He$^4$ \prl   24:715.

\bibitem{colloid}
Ivlev A, L\"owen H, Morfill G and Royall P (2012) {\em Complex Plasmas and Colloidal Dispersions}, (World Scientific, Singapore).
%

\bibitem{tinkham}
Tinkham M (2004) {\em Introduction to Superconductivity, 2$^{nd}$ Ed.} (Dover Publications, New York). 
%
\bibitem{toledano}
Toledano P (2007) Theory of the amorphous solid state: Non-directional elastic vortices and a superhard crystal state {\em Europhys. Lett.} 78:46003.

\bibitem{falk}
Falk ML and Langer JS (1998) Dynamics of viscoplastic deformation in amorphous solids \pre 57:7192.
%
%\bibitem{sas0}
%The Sci. Rep.

\bibitem{sas1}
Ganguly S, Sengupta S, Sollich P, and Rao M (2013) Nonaffine displacements in crystalline solids in the harmonic limit \pre 87:042801.
%
\bibitem{sas2}
Ganguly S, Sengupta S, and Sollich P (2015) Statistics of non-affine defect precursors: tailoring defect densities in colloidal crystals using external fields {\em Soft Matter} 11:4517.
%
\bibitem{sas3}
Mitra A, Ganguly S, Sengupta S, and Sollich P (2015) Non-affine fluctuations and the statistics of defect precursors in the planar honeycomb lattice {\em JSTAT} P06025.

\bibitem{sas4}
Ganguly S, Nath P, Horbach J, Sollich P, Karmakar S and Sengupta S (2017) Equilibrium and dynamic pleating of a crystalline bonded network {\em J. Chem. Phys.} 146:124501.

\bibitem{sas5}
Ganguly S and Sengupta S (2017) Excess vibrational modes of a crystal in an external non-affine field  {\em J. Chem. Sci.} 129:891. 

\bibitem{sas6}
Ganguly S, Mohanty PS, Schurtenberger P, Sengupta S, and Yethiraj A (2017) Contrasting the dynamics of elastic and non-elastic deformations across an experimental colloidal Martensitic transition {\em Soft Matter} 13:4689.

\bibitem{popli}
Popli P, Ganguly S and Sengupta S (2018) Translationally invariant colloidal crystal templates {\em Soft Matter} 14:104.

\bibitem{frenkel}
Frenkel D and Smit B (2002) {\em Understanding Molecular Simulations} (Academic Press, San Diego).

\bibitem{binder}
Binder K and Heermann D (2010) {\em Monte Carlo Simulation in Statistical Physics: An Introduction, 5$^{th}$ Ed.}
(Springer, Berlin).
%
\bibitem{SUS} 
Virnau P and M\"uller M (2004) Calculation of free energy through successive umbrella sampling  {\em J. Chem. Phys.} 120:10925.
%
% 
\bibitem{borgs90}
Borgs C and Koteck\'y R (1990) A rigorous theory of finite-size scaling at first-order phase transitions {\em J. Stat. Phys.} 61:79.
%
\bibitem{ferrenberg88}
Ferrenberg AM and Swendsen RH (1988) New Monte Carlo technique for studying phase transitions  \prl 61:2635.

\bibitem{tfot}
Binder K (1987) Theory of First-Order Phase Transitions {\em Rep. Prog. Phys.} 50:783. 

\bibitem{inter}
Statt A, Virnau P and Binder K (2015) Finite-size Effects on liquid-solid phase coexistence and the estimation of crystal nucleation barriers \prl 114:026101. 
\bibitem{becker}Becker R, D\"oring W (1935) The kinetic treatment of nuclear formation in supersatu- rated vapors. Ann Phys (Weinheim, Ger) 24:719?752.

\bibitem{bruinsma}
Bruinsma R, Halperin BI, Zippelius A (1982) Motion of defects and stress relaxation in two-dimensional crystals  \prb 25:579.
\bibitem{yip}
Bulatov V, Abraham FF, Kubin L, Devincre B, Yip S (1998)
Connecting atomistic and mesoscale simulations of crystal plasticity.
{\em Nature}, 391:669.
\bibitem{allen}
Allen MP and Tildesley DJ (1987) {\em Computer Simulation of Liquids} (Oxford University Press, Oxford).
\bibitem{sc-cnt} 
J. Shillcock and U. Seifert (1998) Escape from a metastable well under a time-ramped force {\em Phys. Rev. E}, {\bf 57}, 7301.

\bibitem{schall}
Schall P, Cohen I, Weitz DA and Spaepen F(2006) Visualizing dislocation nucleation by indenting colloidal crystals. {\em Nature} 440:319.
\bibitem{julie} J. Li, K. J. Van Vilet, T. Zhu, S. Yip, and S. Suresh, Nature (London) {\bf 418}, 307 (2002). 
\bibitem{julie2} J. Li, T. Zhu, S. Yip, K. J. Van Vliet, and S. Suresh, Mat. Sc. and Eng. {\bf A365}, 25 (2004).
\bibitem{dis-nuc}
S. Ryu, K. Kang and W. Cai (2011) Entropic effect on the rate of dislocation nucleation {\em Proc. Natl. Acad. Sc.} {\bf 108}, 5174. 

%
%Glasses
%
%\bibitem{PNAS2}
%Lin J et al. (2014) Scaling description of the yielding transition in soft amorphous solids at zero temperature.
%{\em Proc. Nat. Acad. Sci.} 111:14382.
%\bibitem{schall2}
%Chikkadi V et al. (2011) Long-Range Strain Correlations in Sheared Colloidal Glasses.
%\prl 107:198303.
%\bibitem{schall3}
%Zaccone A, Schall P and Terentjev EM (2014) Microscopic origin of nonlinear nonaffine deformation in bulk metallic glasses. \prb,  90:140203(R).
%\bibitem{manasa}
%Nagamanasa KH et al. (2014) Experimental signatures of a nonequilibrium phase transition governing the yielding of a soft glass. \pre, 89:062308.
% Nonaffine

%
%\bibitem{manning}
%Manning ML and  Liu AJ (2011) Vibrational Modes Identify Soft Spots in a Sheared Disordered Packing.
%\prl,  107:108302. 
%S. S. Schoenholz, A. J. Liu, R. A. Riggleman, and J. Rottler, 
%arXive:1404.1403v1.
%
%\bibitem{denisov}
%Denisov DV et al. (2015) Sharp symmetry-change marks the mechanical failure transition of glasses. {\em Sci. Rep.} 5:14359.
%%


%
%\bibitem{dhar}
%Thomas PB and Dhar D. (1993) J. Phys. A: Math. Gen. 26:3973
%
%\bibitem{yscale}
%Alder, J.F.; Phillips, V.A. (1954) The effect of strain rate and temperature on the resistance of aluminium, copper, and steel to compression. J. Inst. Met. 83: 80-86.
%
\bibitem{HOT} 
Spalding GC, Courtial J and Leonardo RD (2008) in D. L. Andrews Ed., {\em Structured Light and its Applications} (Elsevier, Oxford).


\bibitem{carmen} 
Miguel M-C, Vespignani A, Zapperi S, Weiss J, Grasso J-R  (2001) Intermittent dislocation flow in viscoplastic deformation. {\em Nature}, 410:667.
\bibitem{zaiser} Zaiser M (2006),  Scale invariance in plastic flow of crystalline solids, {\em Advances in Physics}, 55:185.
\bibitem{scaling} Dimiduk DM et al. (2006) Scale-Free Intermittent Flow in Crystal Plasticity. {\em Science} 312:1188.
\bibitem{sethna} Papanikolaou S (2012) Quasi-periodic events in crystal plasticity and the self-organized avalanche oscillator.
{\em Nature}, 490:517.
\bibitem{jamming}Tsekenis G, Goldenfeld N, and Dahmen KA (2011) Determination of the universality class of crystal plasticity, \prl, 106:105501.

\bibitem{debankur}S. Ganguly, D. Das,  Horbach J, Sollich P, Karmakar S and Sengupta S, manuscript in prepraration. 

\bibitem{falk-review}
Falk ML and Langer JS (2010) Deformation and Failure of Amorphous, Solid like Materials.
{\em Annu. Rev. Condens. Matter Phys.} 2:353.

\bibitem{glassbook}
Binder K and Kob W (2011) {\em Glassy Materials and Disordered Solids: An Introduction to Their Statistical Mechanics} (World Scientific, Singapore).

\bibitem{chaudhuri13}
Chaudhuri P and Horbach J (2013)
Onset of flow in a confined colloidal glass under an imposed shear stress.
{\em Phys. Rev. E} 88:040301(R)

\bibitem{barrat11}
Barrat J L and Lema\^itre A (2011), Chapter 8 in
{\em Dynamical Heterogeneities in Glasses, Colloids, and Granular Materials},
edited by Berthier L, Biroli G, Bouchaud J P, Cipelletti L, and van Saarloos W,
(Oxford University Press, Oxford)


%

%\bibitem{szamel2}
%Szamel G and Flenner E (2011) Emergence of Long-Range Correlations and Rigidity at the Dynamic Glass Transition \prl 107:105505.
%
\bibitem{perc}
Shrivastav GP, Chaudhuri P, and Horbach J (2016) Yielding of glass under shear: A directed percolation transition precedes shear-band formation \pre 94:042605.

\bibitem{PNAS2}
Lin J et al. (2014) Scaling description of the yielding transition in soft amorphous solids at zero temperature.
{\em Proc. Nat. Acad. Sci.} 111:14382.
\bibitem{schall2}
Chikkadi V et al. (2011) Long-Range Strain Correlations in Sheared Colloidal Glasses.
\prl 107:198303.
\bibitem{schall3}
Zaccone A, Schall P and Terentjev EM (2014) Microscopic origin of nonlinear nonaffine deformation in bulk metallic glasses. \prb,  90:140203(R).
\bibitem{manasa}
Nagamanasa KH et al. (2014) Experimental signatures of a nonequilibrium phase transition governing the yielding of a soft glass. \pre, 89:062308.
\bibitem{denisov}
Denisov DV et al. (2015) Sharp symmetry-change marks the mechanical failure transition of glasses. {\em Sci. Rep.} 5:14359.
%
\bibitem{itamar}
Parisi G, et al. (2017) Shear bands as manifestation of a criticality in yielding amorphous solids. {\em Proc. Nat. Acad. Sci.} 114:5577?5582
%\bibitem{linde} 
%F. Lindemann, 
%Z. Phys. {\bf 11}, 609, (1910)
%%
%
\bibitem{chaudhuri14}
Chaudhuri P and Horbach J (2014)
Poiseuille flow of soft glasses in narrow channels: From quiescence to steady state.
{\em Phys. Rev. E} 90:040301(R)

\bibitem{shrivastav16}
Shrivastav G P, Chaudhuri P, and Horbach J (2016)
Heterogeneous dynamics during yielding of glasses: Effect of aging.
{\em J. Rheol.} 60:835

\bibitem{sentjabrskaja15}
Sentjabrskaja T, Chaudhuri P, Hermes M, Poon W C K, Horbach J, Egelhaaf S U, and Laurati M (2015)
Creep and flow of glasses: strain response linked to the spatial distribution of dynamical heterogeneities.
{\em Sci. Rep.} 5:11884

\end{thebibliography}
\end{document}